\documentclass[12pt]{article}

\usepackage{amsmath}
\usepackage{amssymb}
\usepackage{amsfonts}
\usepackage{eucal}
\usepackage{epsfig}

\newcommand{\be}{\begin{equation}}
\newcommand{\bea}{\begin{eqnarray}}
\newcommand{\eea}{\end{eqnarray}}
\newcommand{\ba}{\begin{array}}
\newcommand{\ea}{\end{array}}
\newcommand{\ee}{\end{equation}}

\expandafter\ifx\csname mathbbm\endcsname\relax

\else

\fi
\makeatletter \@addtoreset{equation}{section}

\makeatother
\newcommand{\Tr}{{{Tr}}}

\def\cN{{\cal N}}
\def\cD{{\cal D}}

\def\cJ{{\cal J}}

\def\susy{supersymmetry}

\def\opt{operator}
\begin{document}

\begin{titlepage}
\hfill \vbox{
    \halign{#\hfil         \cr
           IPM/P-2005/034   \cr
           hep-th/0505129 \cr
           } 
      }  
\vspace*{6mm}
\begin{center}
{\Large{\bf{ LLL vs. LLM:}}}\\
\vspace*{5mm} {\large {\bf Half BPS Sector of $\cN=4$ SYM $\equiv$
Quantum Hall System}} \vspace*{3mm}
\vspace*{7mm}

 {\bf A. Ghodsi$^{1}$, A. E. Mosaffa$^{1}$, O. Saremi$^{2}$, M. M.
 Sheikh-Jabbari$^{1}$}

\vspace*{0.4cm}

{\it ${}^1${Institute for Studies in Theoretical Physics and Mathematics (IPM)\\
P.O.Box 19395-5531, Tehran, Iran}}

\vspace*{0.4cm}

{\it ${}^2${Physics Department, University of Toronto, Toronto,\\
Ontario, Canada M5S 1A7 }}

\vspace*{0.2cm}

{\tt ahmad,mosaffa, jabbari@theory.ipm.ac.ir,$\,$
omidsar@physics.utoronto.ca}

\vspace*{2cm}
\end{center}

\begin{center}
{\bf\large Abstract}
\end{center}

In this paper we elaborate on the correspondence between the
quantum Hall system with filling factor equal to one and the
$\cN=4$ SYM theory in the 1/2 BPS sector, previously mentioned in
the [hep-th/0409174, 0409115]. We show the equivalence of the two
in various formulations of the quantum Hall physics. We present an
extension of the noncommutative Chern-Simons Matrix theory which
contains independent degrees of freedom (fields) for particles and
quasiholes. The BPS configurations of our model, which is a model
with explicit particle-quasihole symmetry, are in one-to-one
correspondence with the 1/2 BPS states in the $\cN=4$ SYM. Within
our model we shed light on some less clear aspects of the physics
of the $\cN=4$ theory in the 1/2 BPS sector, like the giant
dual-giant symmetry, stability of the giant gravitons, and stringy
exclusion principle and possible implications of the (fractional)
quantum Hall effect for the AdS/CFT correspondence.


\end{titlepage}


\section{Introduction}

In 1973 't Hooft showed that \cite{'tHooft} all correlators of a
$U(N)$ gauge theory, including $\cN=4$ D=4 Supersymmetric
Yang-Mills (SYM) theory, admit a double expansion which can be
arranged in powers of $Ng^2_{YM}$ and $1/N$ where the $1/N$
expansion is encoding the topology of the corresponding  Feynman
diagrams, as we have in string theory. This observation found a
full realization within the celebrated AdS/CFT framework
\cite{MAGOO}.

In the $\cN=4$ $U(N)$ SYM, there are specific sectors, i.e. set of
\opt s, in which the above double expansion reduces to a single
expansion in powers of $1/N$. The \opt s of this sector are chiral
primaries (and their descendants) whose scaling dimension is exact
and the $g_{YM}$ corrections are absent due to supersymmetry. In
fact the large amount of \susy\ removes any $g_{YM}$, perturbative
or non-perturbative, dependence in all the $n$-point functions of
chiral primaries. The chiral primary \opt s preserve 16
supersymmetries of the superconformal algebra of the $\cN=4$ SYM
theory, $PSU(2,2|4)$, and hence they all belong to the 1/2 BPS
sector of the theory. As such 1/2 BPS sector provides us with a
laboratory to concentrate on the 1/N behavior and the
combinatorics of the $n$-point functions.

In the 1/2 BPS sector, which will be our main focus in this paper,
the $\cN=4$ $U(N)$ SYM simplifies significantly and essentially
becomes equivalent to a system of $N$ 2d fermions \cite{Beren1, Antal}.
These fermions are living in a specific sector of a 2d harmonic
oscillator potential. The 1/2 BPS condition restricts the dynamics
of the system further down to a one dimensional $N$ fermion
system, with degenerate energy levels. These facts will be
reviewed and explained further in section \ref{sec3.1}.

One would try to explore the fermionic nature appearing in the
analysis of the 1/2 BPS sector in the dual gravity picture. This
has been carried out in a novel work by Lin-Lunin-Maldacena (LLM)
\cite{LLM}. LLM constructed type IIB supergravity solutions
compatible with the supersymmetries preserved in the 1/2 BPS
sector of $\cN=4$ SYM; i.e. LLM found {\it static, non-singular}
deformations of $AdS_5\times S^5$ geometry preserving at least 16
supercharges with $SO(4)\times SO(4)\times U(1)\subset
SO(4,2)\times SO(6)$ isometries, the $U(1)$ corresponding to
translations along the globally defined time-like (or in special
cases light-like) Killing vector. The solutions of LLM are only
determined through a  single function $z$, which is a function of
three coordinates usually denoted by $(x_1,x_2)$ and a
non-negative coordinate $y$, and $z/y^2$ satisfy a six dimensional
Laplace equation \cite{LLM}. The LLM geometries are then
completely specified by giving the value of the function $z$ at
$y=0$. The smoothness condition forces $z_0(x_1,x_2)\equiv
z(x_1,x_2, y=0)$ to take only values $+1/2$ or $-1/2$, a very
restrictive and strong condition. Hence, LLM made a one-to-one
correspondence between the 1/2 BPS sectors of $AdS_5\times S^5$
deformations and the fermionic description of the $\cN=4$ SYM, by
directly identifying $(x_1,x_2)$ plane with the phase space of the
one dimensional fermions mentioned earlier. Explicitly this
identification implies that%
\be\label{x1x2-NC}%
[x_1,x_2]=2\pi i\ l_p^4\ , %
\ee%
where $l_p$ is the ten dimensional Planck length. That is, the
$(x_1,x_2)$ plane in the geometry turns out to be a noncommutative
Moyal plane, a result coming from and confirmed by the quantum
gravity considerations (resulting via AdS/CFT from the $\cN=4$ SYM
description). To be precise and to make the above correspondence
exact, LLM borrowed one fact from the semiclassical physics,
namely quantization of the supergravity fiveform flux. The
smoothness condition $z_0=\pm 1/2$, is then closely related to the
Pauli exclusion principle in the fermion picture. LLM used a
convenient color coding: denote the $z_0=-1/2$ with black and
$z_0=+1/2$ with white. In the fermion picture, $z_0-1/2$ is the
fermion density (in its phase space), i.e. black regions are
filled with fermions and white regions are empty.

The system of 2d fermions moving in a constant background magnetic
field has an interesting sector in which energy levels of fermions
are degenerate and fermions are labeled by their angular momentum
quantum number, the lowest Landau level (LLL). In the LLL the
phase space of fermions is equivalent to the one we encountered in
the 1/2 BPS sector of $\cN=4$ SYM. On the other hand (almost) all
the interesting physics of the Quantum Hall Effect (QHE) can be
described in terms of LLL's. Therefore, one should be able to
build a dictionary between $\cN=4$ SYM 1/2 BPS \opt s (or LLM
geometries) and the quantum Hall physics, within which one can
learn more about the $\cN=4$ SYM and hence quantum gravity from
QHE. This is indeed what we are aiming for in this paper. Some
preliminary steps in this direction have been taken in
\cite{Beren2}.

To build the QHE/SYM correspondence we need to review, very
briefly, the standard descriptions of QHE. This is done in section
\ref{sec2}, where we present three different formulations for
studying quantum Hall systems. In section \ref{sec3}, we show how
the 1/2 BPS sector of SYM is directly mapped into either of the
quantum Hall descriptions. We show this at the level of the
actions and Hilbert spaces. As has also been mentioned in
\cite{Beren2}, we will see $\cN=4$ SYM is a specific quantum Hall
system with filling factor $\nu=1$. We then focus on the
quasiparticles of the corresponding quantum Hall system and argue
that they are mapped to giant gravitons \cite{S-giants,
AdS-giants} in the gravity (gauge theory) picture.

Quantum Hall system with $\nu=1$ has a particle/quasihole
symmetry. In section \ref{sec4} we elaborate further on this
issue. There we present an extension of the Chern-Simons matrix
model action by promoting the quasiholes to new degrees of
freedom, new fields. That is we present an effective field theory
consisting of two fields, one for particles and one for
quasiholes. The BPS solitons of our proposed action exhibit
particle/quasihole symmetry. In section \ref{sec4} we present
several solutions of these BPS equations and show that our BPS
configurations are in one-to-one correspondence with chiral
primary \opt s of $\cN=4$ SYM. Using our proposed action
 we study stability of its BPS solitons and show that classically
 there is no transition between giant graviton states.
 The last section
is devoted to discussions on our results and further extensions
and generalizations of QHE/SYM correspondence.

\section{Lightening Review of the Quantum Hall System}\label{sec2}

Integer Quantum Hall effect (IQHE) is simply quantization of the
Hall conductance in $e^2/h$ units as $\sigma_{xy}=\nu e^2/h$,
where $\nu$ is an integer quantum number. This phenomenon is
exhibited by condensed matter systems which can be approximated as
an effectively two dimensional ideal electron gas living in a
strong magnetic field. Fractional values of the quantum number
$\nu$ has also been observed. The corresponding Hall effect is
called Fractional Quantum Hall effect (FQHE). Fractional and
integer quantum hall effects have two quite different underlying
physics. Physics of the FQHE involves strong correlations among
the electrons. Collective excitations of the electron gas in the
fractional case have fractional charges and statistics which is
something between ordinary Bose and Fermi statistics. Quantum
description of the Hall system for $\nu^{-1}\in {\mathbb{Z}}$ is
given by the so-called Laughlin wavefunction. Laughlin wave
function encodes the edge fluctuations of an incompressible
gapless fluid. There also exists an algebraic approach to the Hall
problem. The algebra of the area preserving diffeomorphisms in two
dimensions ($w_{\infty}$ algebra) has been studied before e.g. see
\cite{w-infty}. The quantum version of this algebra so called
$W_{1+\infty}$ describes the edge excitations of the Hall droplet.
For more details refer to \cite{W-infty, Karabali}.

There have been three different approaches to quantum Hall system
in the literature. The more standard one is based on quantum
mechanics of some non-relativistic fermions in an (strong)
external magnetic field in the Lowest Landau Level (LLL) (for
review see \cite{Girvin}) and second one is the effective field
theory description in terms of (noncommutative) Chern-Simons gauge
theory \cite{Susskind}. The third one is the Matrix Chern-Simons
theory which interpolates between the field theory description and
the quantum mechanical one. In this section we briefly review these
three approaches.

\subsection{Landau problem and Quantum Hall effect}\label{sec2.1}

Here we study the Landau problem in two different bases. This
would  enable us to draw an explicit connection between the Landau
problem and the half BPS sector of the $\cal N =$4 SYM.

Let us start with a single non-relativistic charged particle
moving in a two dimensional plane transverse to a constant
magnetic field i.e., the Landau problem. The Hamiltonian for the
system
is%
\begin{equation} \label{Landau-Ham}%
 H=\frac{1}{2m}(p_i-\frac{eB}{2c}\epsilon_{ij} x_j)^2\ ,
\end{equation}%
where $i,j=1,2$ and $x_i, p_i$ are the corresponding phase space
conjugate variables and $B$ is the strength of the magnetic field.
Next let $\Pi_i=p_i-\frac{eB}{2c}\epsilon_{ij} x_j$, it is then
readily seen that
\begin{equation}
[\Pi_1, \Pi_2]=-i\hbar \frac{eB}{c}\ ,
\end{equation}
and hence if we call $Y_1=\frac{c}{eB} \Pi_2$,
$[Y_1,\Pi_1]=i\hbar$, the Hamiltonian takes the form
\begin{equation}
H=\frac{1}{2m}\left(\Pi_1^2+ (\frac{eB}{c})^2 Y_1^2\right)\ ,
\end{equation}
which is the Hamiltonian for a simple one dimensional Harmonic
oscillator with frequency%
 \begin{equation}\label{omegaB}%
  \omega_0=\frac{eB}{mc}\ .%
\end{equation}
The spectrum of the Hamiltonian is%
\begin{equation}\label{spectrum}%
E=\hbar\omega_0(n+\frac{1}{2})\ .
\end{equation}%
Noting that $\cal J$$ = \epsilon_{ij} x_i p_j$ also commutes with
the Hamiltonian the energy eigenstates are infinitely degenerate
and their degeneracy is labeled by $J$, eigenvalues of $\cal J$,
which can be any arbitrary integer.

Now, let us analyze the above Hamiltonian in another way. Expand
the square to obtain %
\begin{eqnarray} \label{2d-oscil}
  H&=&\frac{1}{2m}p^2+ \frac{e^2B^2}{8mc^2} x^2+
  \frac{eB}{2mc} \epsilon_{ij}x_ip_j\ .
\end{eqnarray}%
The first two terms of the above Hamiltonian is the Hamiltonian
for a two dimensional harmonic oscillator, $H_0$, with frequency
$\omega_0/2$. The spectrum of this part is
\begin{equation}
E_0=\frac{1}{2}\hbar\omega_0(n+n'+1)\ .
\end{equation}
The next part is proportional to the angular momentum which has
the spectrum $J=\hbar(n'-n)$, where both $n$ and $n'$ are
non-negative integers. Putting these two contributions together we
obtain the spectrum of the whole Hamiltonian $H$ to be exactly
given by \eqref{spectrum}.

The lowest energy state, the lowest Landau level (LLL), is then
given by $n=0$ but arbitrary $n'$. In the lowest Landau level,
ignoring the zero point energy, $H_0$ is proportional to $J$, i.e.%
 \begin{equation}
H_0-\frac{1}{2}\hbar\omega_0= \frac{1}{2}\hbar\omega_0 J\ .
\end{equation}%
In the lowest Landau level, which is known to describe the quantum
Hall physics, the Hamiltonian is essentially $J$, or the action
corresponding to the system is 
\begin{equation}\label{Susskind-action}%
S=\frac{eB}{2c}\int dt \epsilon_{ij}x_i\dot x_j\ .%
\end{equation}%

\subsection{Fluid description of Quantum Hall
effect}\label{sec2.2}

Consider a system of finite number of non-relativistic interacting
particles. A continuum (field theory) description of this fluid is
obtained by promoting the particle labels to a continuous
co-moving coordinate $y$. The real space density is given by
\begin{equation}\label{rho-density}
\rho=|\frac{\partial y}{\partial x}|\rho_{0}\ ,
\end{equation}%
where $x$ is the real space position.
It can be seen that if the fluid is incompressible the
corresponding continuum Lagrangian has a gauge invariance under
area preserving diffeomorphisms (APD) in the $y$ plane
\cite{Susskind}. Small fluctuations around the static background
solution $x_{i}=y_{i}$ is closely connected to Electrodynamics if
we parameterize these fluctuations as
\begin{equation}
x_{i}=y_{i}+\epsilon_{ij}\frac{A_{j}}{2\pi\rho_0}\ ,
\end{equation}%
where $A$ plays the role of electromagnetic vector potential.

Now consider a fluid with charged particles moving in a constant
magnetic field. The Lagrangian acquires a new term induced by the
background magnetic field. The APD's would still keep the new
action invariant and \eqref{rho-density} can be rewritten as
\begin{equation}\label{Poisson-B}
1=\frac{1}{2}\epsilon_{ij}\epsilon_{ab}\frac{\partial
x_{b}}{\partial y_{j}}\frac{\partial x_{a}}{\partial y_{i}}\equiv
\frac{1}{2}\epsilon_{ab}\{x^a,x^b\}_{P.B.}\ .
\end{equation}%
In the strong magnetic field limit the action is basically
dominated by the magnetic field term (setting $c=1$), i.e.
\[
S=\frac{eB\rho_0}{2}\int dtd^2y \epsilon_{ab}x^a{\dot x}^b\ .
\]
(Dropping other terms in the action is equivalent to restricting
to lowest Landau level.) In this limit the equation of motion and
the constraint can be encapsulated in a single action via
introducing a non-dynamical time component of $A$, $A_0$:
\begin{equation}\label{ContinuumAction}
L=\frac{eB\rho_{0}}{2}\epsilon_{ab}\int d^2y[(\dot
X_{a}-\frac{1}{2\pi\rho_{0}}\{X_{a},A_{0}\})X_{b}+\frac{\epsilon_{ab}}{2\pi\rho_{0}}A_{0}]\ ,
\end{equation}
where the bracket is the Poisson bracket defined in
\eqref{Poisson-B}.

This theory admits vortex solutions. Chern-Simons vortex is
basically radial disturbance of the fluid toward or away from the
center of the vortex proportional to the $q/r$ where $q$ is
related to the excess or deficit charge of the vortex by
\begin{equation}
e_{qp}=\rho_{0}qe\ ,
\end{equation}%
and $r$ measures the distance from the center of the vortex. These
are quasihole or quasiparticle states of the Hall fluid in the
continuum description. Semiclassical quantization of this theory
implies that \cite{Susskind}
\begin{equation}
e_{qp}=2\pi\frac{\rho_{0}}{B}=\nu e\ ,
\end{equation}
where $\nu=\frac{2\pi\rho_0}{eB}$ is the filling fraction. Filling
fraction is the ratio of the number of the electrons to the
magnetic flux (that is, inverse of magnetic flux per particle).
Full quantization of the theory gives rise to quantization of the
$\nu$ inverse. Filling fraction also controls statistics of the
collective excitations of the fluid.

\subsection{NC Chern-Simons Matrix model description of
QHE}\label{sec2.3}

APD's (or $w_\infty$ algebra), which reflect symmetry of the
system under relabeling of the particles in an incompressible
fluid, translate into the gauge symmetries of the continuum
theory. A gauge theory based on this gauge invariance would be
able to capture some long distance physics but it is unable of
incorporating the intrinsic granular structure of the fluid. It
turns out that a description of the system which is more faithful
to the underlying discrete physics is a matrix model description
of the fluid. In this description classical configuration of the
$N$ number of electrons is replaced by the space of $N\times N$
Hermitian matrices. The action \eqref{ContinuumAction} can be
generalized to a matrix theory \cite{Susskind}
\begin{equation}\label{MatrixLagrangian}
L=\frac{eB}{2}\epsilon_{ab}Tr(\dot
X_{a}-i[X_{a},A_{0}])X_{b}+eB\theta Tr A_{0}\ ,
\end{equation}
where $\theta=1/(2\pi\rho_{0})$ plays the role of the
noncommutativity parameter. In this action the APD's are replaced
by the $U(N)$ gauge symmetry. Constraint equation is obtained by
varying this action with respect to $A_{0}$
\begin{equation}\label{Class-NC}
[X_{a},X_{b}]=i\theta\epsilon_{ab} \ .
\end{equation}%
Although we started with a finite $N$, due to antisymmetric nature
of the commutator,\eqref{Class-NC} can only be solved for infinite
size matrices. Therefore, the model describes ``infinite'' number
of particles. We will return to this point momentarily.

It is worth emphasizing that there are two different kinds of
noncommutativity not to be confused with each other. One of them
is controlled by $1/(eB)$ and has a quantum mechanical origin;
$X_{1}$ and $X_{2}$ are canonically conjugate and hence do not
commute as quantum operators, i.e.
\[
[\hat{X}_1, \hat{X}_2]_{opt.}=\frac{i}{eB}\ .
\]%
The second one is encoding the APD invariance of the theory (in
the continuum case) or its permutation subgroup (in the discrete
case i.e. the matrix model description). The latter
noncommutativity is controlled by $\theta=1/(2\pi\rho_{0})$
\[
[X_1,X_2]_{Mat.}=i\theta\ .
\]%

Statistics of the Chern-Simons particles is determined by the
filling fraction $\nu$. The ratio of the two noncommutativities,
$\nu^{-1}=eB\theta$ is ought to be an integer if we demand the
action \eqref{MatrixLagrangian} to be invariant under large gauge
transformations \cite{level-quantization}. It can be shown that
there is a density-statistics connection; depending on
$\nu=1/(2n+1)$ or $\nu=1/(2n)$ for integer $n$, the Hall fluid
excitations are either Fermions or Bosons.

At this point it is instructive to pause for a moment and make a
comparison with the $\cal N$=4 $SYM$ theory in its half BPS
sector. It is rather clear that the above mentioned matrix model
cannot have an origin as a particular sector of a ``finite'' $N$
($U(N)$ is the gauge group) SYM theory. It is impossible to
satisfy the APD invariance constraint by means of finite
dimensional matrices as of finite $N$ $\cal N$=4 SYM theory. As it
was explained earlier the same problem occurs if one is to write a
matrix model for a Hall system with finite number of particles.

Matrix model for Hall systems with finite number of particles has
been discussed in \cite{Polychronakos}. Finite $N$ matrix models
can be constructed by introducing new degrees of freedom so-called
edge states. In the presence of the edge state commutator gets
modified. This anomaly provides the opportunity to satisfy the APD
invariance constraint by finite $N$ matrices. Now we proceed to
review some general aspects of the finite $N$ Chern-Simons matrix
model.

\subsection{Finite dimensional NC Chern-Simons Matrix
models}\label{sec2.4}

The starting point is to modify the action
\eqref{MatrixLagrangian} with extra degrees of freedom called edge
state \cite{Polychronakos}
\begin{equation}\label{Polykronachos-action}
L=\frac{B}{2}Tr{\epsilon_{ab}(\dot X_{a}+i[A_{0},X_{a}])X_{b}+
B\theta A_{0}}+B \Psi^{\dagger}(i\dot
\Psi-A_{0}\Psi)-\frac{1}{2}\omega^2 (X_a)^2,
\end{equation}%
where $\Psi$, the edge state, is a complex valued vector field in
the fundamental of the $U(N)$ gauge group and we have set $e=1$.
The $X^2$ term has been added to make a droplet the lowest energy
state \cite{Polychronakos} and setting $\omega=0$
\eqref{Polykronachos-action} reduces to NC Chern-Simons Matrix
model. The constraint gets modified as follows
\begin{equation}\label{ModifiedConstraint}
-i[X_{1},X_{2}]+\Psi\Psi^{\dagger}-\theta=0\ .
\end{equation}
Taking the trace would not lead to any inconsistency instead it
gives
\begin{equation}
\Psi^{\dagger}\Psi=N\theta\ .
\end{equation}
Using equation of motion for $\Psi$ in temporal gauge, one obtains
\begin{equation}\label{Psi-soln}
\Psi=\sqrt{\theta N}|v\rangle\ ,
\end{equation}
where $|v\rangle$ is an arbitrary constant unit vector. The
constraint \eqref{ModifiedConstraint} now reads
\begin{equation}
[A,A^{\dagger}]=2\theta(1-N|v\rangle\langle v|)\ ,
\end{equation}
where
\begin{equation}
A=X_{1}+iX_{2}\ .
\end{equation}
There are various finite dimensional solutions to the above
constraint equation corresponding to different Hall states. For
instance
\begin{eqnarray}
A&=&\sum_{n=0}^{N-1}\sqrt{2n\theta}|n-1 \rangle \langle n|\ , \\
|v \rangle &=& |N-1 \rangle\label{GaugeChoice}\ ,
\end{eqnarray}
where \eqref{GaugeChoice} is only a freedom in gauge choice
originating from the time independent part of the $U(N)$ gauge
invariance of the theory, is a solution representing a circular
quantum Hall droplet of radius $\sqrt{2N\theta}$. $|m\rangle$ is a
harmonic oscillator basis. The radius-squared matrix coordinate $
R^2=X_{1}^2+X_{2}^2=A^{\dagger}A $ is diagonal in the harmonic
basis and its highest eigenvalue goes like $N\theta$ which
suggests that we are talking about a finite size quantum Hall
state or Hall {\it droplet} \cite{Polychronakos}. There are other
types of excitations which will be of our interest later on:
quasihole states. Quasihole states with charge $-q$ at the origin
satisfying \eqref{ModifiedConstraint} can be constructed as well
\begin{equation}\label{droplet+hole}
A=\sqrt{2\theta}(\sqrt{q}|N-1 \rangle\langle 0|+\sum_{n=1}^{N-1}\sqrt{n+q}|n-1 \rangle \langle n|)\ . \\
\end{equation}
Looking at $R^2$ eigenvalues reveals that the lowest eigenvalue is
proportional to $2\theta q$. This means that there is a circular
hole of area proportional to $2\theta q$ at the origin. Of course
the radius of the droplet itself has also changed to take care of
the total number of the particles inside the droplet which is
fixed to be $N$. There are no quasiparticle excitations
(accumulation of the particles) in this model.

\section{Building the  SYM/Quantum Hall  Dictionary}\label{sec3}

In the previous section we reviewed various approaches to quantum
Hall problem and made connections between them. In this section we
show that how $\cN=4$ $U(N)$ SYM theory, in the 1/2 BPS sector, is
related to quantum Hall problem and each of the above approaches.
This will be done first at the level of the corresponding actions
and then by relating the 1/2 BPS SYM operators and the quantum
Hall states. We also briefly discuss the gravity picture via
AdS/CFT duality.

\subsection{Effective action in the 1/2 BPS sector}\label{sec3.1}

Let us start with the $\cN=4$ $U(N)$ SYM action on $R\times S^3$
and denote one of the three complex scalars present in the $\cN=4$
gauge multiplet by $Z$. In the 1/2 BPS sector the \opt s can only
be made out of $Z$ and moreover these \opt s cannot have
non-trivial dependence on the $S^3$. In the 1/2 BPS sector we
should preserve $SO(4)\times SO(4)\times U(1)\subset SO(4,2)\times
SO(6)_R$ of the theory and are only allowed to perturb the theory
by chiral primary \opt s, i.e. \opt s only made out of $Z$ and not
$Z^\dagger$ or any other fields \cite{Beren1, Antal}. Therefore,
the action relevant to this sector is simply%
 \be\label{reduced-action}
S_{reduced}=\frac{1}{2}\int dt\ Tr\left( ( D_0 Z)^\dagger D_0 Z-
Z^ \dagger Z-\frac{1}{2}[Z,Z^\dagger]^2\right) ,
\ee%
where we have used the conformal invariance of the theory and
chosen the radius of the $S^3$ such that there are no prefactors
in the action and  we have also rescaled $t, Z, A_0$ such that
they are all dimensionless. The above action, ignoring the last
term, is the action for $N^2$ uncoupled two dimensional harmonic
oscillators with frequency one. One should, however, remember that
not all the elements of the $N\times N$ matrices are independent
and dynamical, as they may be related by the $U(N)$ gauge
transformations. We will come back to the issue of gauge fixing
later on in this section.

The Dilatation \opt\ in the  sector containing \opt s only made
out of $Z$ and $Z^\dagger$, in the first lowest order in
$g_{YM}^2$ is of the form:\footnote{It is worth noting that the
D-term $[Z,Z^\dagger]^2$ in the action \eqref{reduced-action} does
not contribute to the dilatation operator.}%
 \be\label{dilatation-opt}
\cD= \Tr\left( Z\frac{\delta}{\delta Z}\right) +\Tr\left(
Z^\dagger\frac{\delta}{\delta Z^\dagger}\right)\ . %
 \ee%
In the same sector the R-charge $J$ is measured by%
 \be\label{Rcharge-opt}
\cJ= \Tr\left( Z\frac{\delta}{\delta Z}\right) -\Tr\left(
Z^\dagger\frac{\delta}{\delta Z^\dagger}\right)\ . %
 \ee%

Note that \eqref{dilatation-opt} is nothing but the Hamiltonian
for $N^2$ 2d harmonic oscillators with the same frequency.
In the 1/2 BPS sector $\cD-\cJ=0$ and hence in this sector%
\be\label{halfBPS-dilat}%
 \cD= \Tr\left( Z\frac{\delta}{\delta Z}\right)\ .
\ee%
(In this sector we are dealing with the \opt s which are only made
out of $Z$.) In the 1/2 BPS sector there is no $g_{YM}$ dependence
in the scaling dimensions of the operators and hence the
dilatation \opt\ \eqref{dilatation-opt} is exact. As it is
manifest in the 1/2 BPS sector we are only left with an
effectively one dimensional system out of the 2d harmonic
oscillator system we started with.

$\frac{\delta}{\delta Z}$ is the momentum conjugate to $Z$,
$\Pi_{Z}$. On the other hand from the $\cN=4$ SYM action, in the
temporal gauge, we have $\Pi_{Z}=\dot Z^\dagger$. Therefore, the
action for a system with the Hamiltonian $\cD-\cJ(=0)$ in the BPS
sector is
simply %
 \be\label{1/2BPS-action}
S_{1/2\ BPS} = \frac{i}{2}\int dt\ Tr \left(Z^\dagger \dot Z-(\dot Z)^\dagger Z\right)\ .%
 \ee %

 It is very instructive to compare the
above discussions and formulae with those of sections 2.1 and 2.4.
In terms of quantities in section 2.1, it
is explicitly seen that the two are related by%
 \be\label{correspondence}
 \begin{split}
 \cD&\leftrightarrow H_0-\hbar\omega_0/2 \cr
 \cJ&\leftrightarrow \cJ\cr
 \cD-\cJ&\leftrightarrow H
 \end{split}
 \ee%
Hence $\cD-\cJ$ is the Hamiltonian of $N$ fermions (see the
arguments below) in the external magnetic field and it is seen
that going to LLL corresponds to taking the BPS states for which
$\cD-\cJ$ vanishes.

The action \eqref{1/2BPS-action}, with $N\times N$ matrices, is
equivalent to the Polychronakos finite $N$ matrix Chern-Simons
theory discussed in section 2.4 after fixing the temporal gauge
$A_0=0$ and replacing the edge state $\Psi$ with the specific
classical solution given in \eqref{Psi-soln}. For the equivalence,
however, one should still  impose the constraint
\eqref{ModifiedConstraint}.

So far we have established the relation between the quantum Hall
system and the 1/2 BPS sector of $\cN=4$ SYM at classical level,
i.e. at the level of the actions. As the next step we would like
to push this further to the level of partition function.
Explicitly, we want to show that the partition function of the
$\cN=4$ $U(N)$ SYM in the 1/2 BPS sector is producing the
``Laughlin'' wavefunction \cite{Girvin}, that is%
\be\label{partition-Laughlin}%
Z_{1/2\ BPS}=\int\ \left(DZD{\bar Z}|_{1/2\ BPS}\right)\
e^{-S_{1/2\
BPS}}=\langle \psi_L|\psi_L\rangle\ , %
\ee%
with%
\be\label{Fermionic-w.f}%
\psi_L=\prod_{i>j=1}^{N} (z_i-z_j)\ e^{-\sum_{i=1}^N {\bar z}_i
z_i/2}\ , %
\ee%
where $z_i,{\bar z}_i$ $i=1,2,\cdots, N$ are the eigenvalues of
the $N\times N$ matrices of $U(N)$ $\cN=4$ SYM, $Z,\ {\bar Z}$. To
show this we need to use the $U(N)$ gauge symmetry to diagonalize
$Z$. We fix the temporal gauge $A_0=0$ and use the remaining time
independent (global) gauge transformations  to diagonalize $Z$.
This, however, is not possible with a single $U(N)$, as $Z$ is a
complex (non-hermitian) $N\times N$ matrix. In order to
diagonalize $Z$ we need to double $U(N)$ to $U(N)\times U(N)$ or
to complexify $U(N)$ to $U(N,\mathbb{C})$. Since we are going to
reduce the computations to 1/2 BPS sector indeed we can use this
extended gauge group. To see this it is more convenient to use the
Hamiltonian path integral with measure $DZ D{\bar Z} D\Pi_Z
D\Pi_{\bar Z}$. In the 1/2 BPS sector, however, $\Pi_Z=i{\bar Z}$
and $\Pi_{\bar Z}=-iZ$. This means that in order to do
computations with the path integral in the 1/2 BPS sector in the
process of the gauge fixing we need to divide the measure by
$Vol_{U(N)}\times Vol_{U(N)}$. This justifies effective extension
of the gauge group needed for diagonalizing $Z$. The rest of the
computations are the standard Van der Monde determinant techniques
\cite{Brezin} leading to%
\[
DZD{\bar Z}|_{1/2\ BPS}=\prod_{i>j=1}^N(z_i-z_j)
\prod_{i>j=1}^N({\bar z}_i-{\bar z}_j)\prod_{i=1}^N dz_id{\bar
z}_i \ .
\]%

The Lagrangian in the 1/2 BPS sector, for diagonalized $Z$ simply
reduces to $\sum_{i=1}^N {\bar z}_i z_i$. This proves the
statements made in \eqref{partition-Laughlin},
\eqref{Fermionic-w.f}.

As the first outcome of the above discussion, we note that the 1/2
BPS sector of $\cN=4$ SYM is equivalent to a Laughlin wavefunction
with $\nu=1$.\footnote{The Laughlin wavefunction for a quantum
Hall system with filling fraction $\nu$ is \cite{Girvin}%
\[
\psi_{Laughlin}=\prod_{i>j=1}^N(z_i-z_j)^{\frac{1}{\nu}}\
e^{-\sum_{i=1}^N {\bar z}_i z_i/2}\ .
\]
} This observation has also been made in \cite{Beren2}. For
$\nu=1$ the wavefunction is antisymmetric with respect to exchange
of any two $z_i, z_j$ and hence the eigenvalues of $Z$ are
describing positions of $N$ fermions in the lowest Landau level
(LLL).

Next, we note the important property of the wavefunctions of the
system of $N$ particles in the LLL: the wavefunction can be
written as%
\[
\psi_{LLL}=f(z_i)\ e^{-\sum_{i=1}^N {\bar z}_i z_i/2}\ ,
\]
where $f$, regardless of the statistics of the underlying
particles and its symmetry behavior under exchange of $z_i$'s,
is a {\it holomorphic} function of $z$'s \cite{Girvin}. This
holomorphicity is then directly related to the fact that the
chiral primaries (1/2 BPS \opt s of $\cN=4$ SYM) are holomorphic
in $Z$.

Finally, as reviewed in section \ref{sec2.1}, in the Landau
problem the coordinates of the 2d particles become noncommutative
\cite{Girvin} and in the LLL particles essentially become one
dimensional. In other words the $(z_i,{\bar z_i})$ plane becomes
the phase space of the particles \cite{Girvin, Beren1} and hence
(in proper units) $[z_i,{\bar z}_j]=\delta_{ij}$. As reviewed
briefly in the introduction, this has become manifest in the LLM
setup \cite{LLM}.

The emergence of an integer Hall system from the 1/2 BPS sector of
SYM finds a simple interpretation within the geometric description
of this sector via LLM. As reviewed in the introduction, the
smoothness condition for these geometries allows for only two
boundary conditions for the function $z_0$ on the noncommutative
plane $(z_i,{\bar z_i})$. In terms of LLM's color coding, this
means that the minimal area accessible to the black regions is the
same as the one for the white regions. Alternatively one can say
that the absence of a minimal black spot corresponds to the
presence of a minimal white spot. On the other hand, for a QHS
with $\nu\equiv 1/k$ (with integer $k$), the minimal area accessible to a particle is
$k$ times as that for a hole. As a result, the density of
particles can acquire $k+1$ different values or alternatively the
absence of a particle is equivalent to the presence of $k$ holes.
The two pictures can thus be related to one another only if $k=1$.

Therefore, $\cN=4$ SYM in the 1/2 BPS sector is describing the
same physics as a quantum Hall system with $\nu=1$ and with a specific
edge state. In what follows we elaborate further on the connection
and relation of the two systems.

\subsection{QH solutions Vs. SYM operators}\label{sec3.2}

In this section we explore the correspondence between the 1/2 BPS
operators of $\cN=4$ SYM and the physical states of an integer
QHS. Dealing with a $U(N)$ gauge theory with definite $N$ calls
for a QHS with a finite number of particles
which is provided by the Polychronakos' construction
\cite{Polychronakos}.

Let us first review the physical states of a finite QHS with
arbitrary $\nu$ (we will later focus on $\nu=1$).
The physical states of this system can be found by quantizing
the corresponding matrix
model \eqref{Polykronachos-action} which results in the quantum
Calogero model with the following Hamiltonian (we take
$\omega=B=1$) \cite{Calogero,
Polychronakos, Calo-Poly}%
\be H=\frac{1}{2}\sum_{n=1}^N(p_n^2+x_n^2)+ \sum_{n\neq
m}\frac{k(k-1)}{(x_n-x_m)^2}\;, \label{calo}%
\ee%
 where $\nu=1/k$
(note that for $k=1$ the second sum vanishes). The eigenstates of
this system are well known and are given by $N$ positive integers
$(f_1,f_2,...,f_N)$ (known as quasinumbers) such that
$f_i+1>f_{i+1}+k$. The ground state of the QHS, $|0\rangle_{QH,k}$,
is given by $f_i=k(N-i)$ and its energy will be $\frac{1}{2}(k
N^2+N(1-k))$. For future use we express the states in terms of
their excitations above the ground state by the nonnegative
integers $(r_1,r_2,\cdots ,r_N)$ where $r_i=f_i-k(N-i)$ such that
$r_1\ge r_2\ge \cdots\ge r_N\ge0$. As explained in
\cite{Polychronakos}, these states describe $N$ independent
harmonic oscillators with an enhanced exclusion principle such
that any two occupied states can not be closer than $k$. For
$k=1$, the particles will thus be ordinary fermions.

One can, on the other hand, find the states of $\cN=4$ SYM in the
1/2 BPS sector. In \cite{Beren1, Antal}, these states have been found by
quantizing a one matrix model with a harmonic oscillator potential
with unit frequency. As described in this paper, different gauge
fixings for the model result in different bases for the spectrum.
The first one, trace basis, consists of $N^2$ free bosonic
harmonic oscillators subject to the constraint that the states
must be neutral under gauge transformations. These states are
expressed in terms of a set of positive integers
$(c_1,c_2,\cdots)$ such that $N\ge c_1\ge c_2\ge\cdots$. Each
number $n$ in this set is attributed to a creation operator
$\beta_n^\dagger$ which acts on the gauge invariant vacuum
$|0\rangle_{tr}$. The ground state energy for the vacuum will thus
be $N^2/2$. The upper bound on $c_i$ comes from the fact that
operators with $n> N$ are not independent.

\begin{figure}[ht]
\begin{center}
\epsfig{file=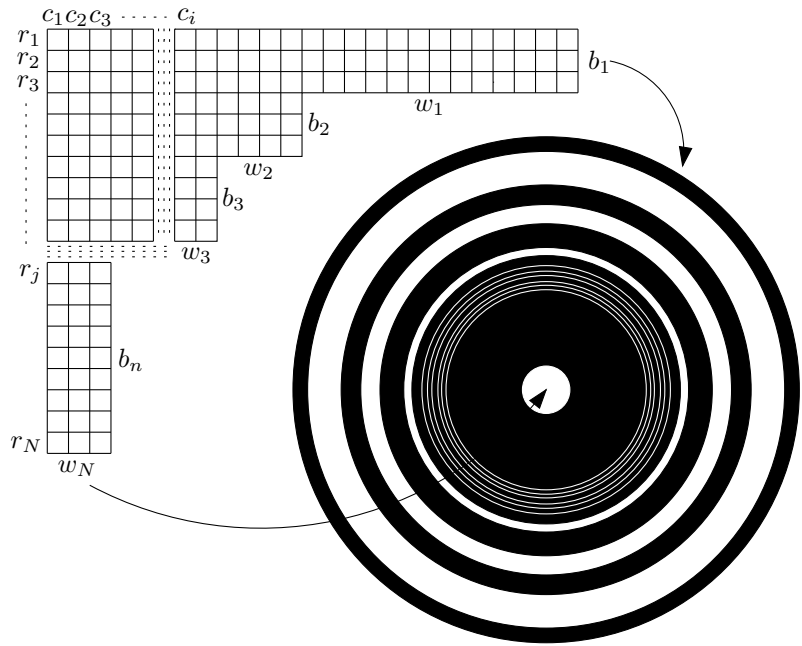} \caption{A generic Young tableau is in
one-to-one correspondence with an LLM configuration of black and
white circular rings, which is also equivalent to a similar
configuration of quantum Hall droplets. There are various ways to
label a Young tableau, each suitable for a different
interpretation in the quantum Hall system. For example one can
identify a tableau by the length of its rows ($r_i$'s) or length
of its columns ($c_i$'s) or  the Dynkin labels,
$(b_1,w_1;b_2,w_2;b_3,w_3;...;b_n,w_n)$ ($b_i$ is counting the
number of zeros in the standard Dynkin labels). The latter
directly correspond to the area of black and white regions in the
quantum Hall droplet picture. The $r_i$'s and $c_i$'s are related
by bosonization of 2d fermion system.}\label{Dynkin-figure}
\end{center}
\end{figure}

A second gauge choice, eigenvalue basis, leads to a system of $N$
free fermionic oscillators. The states are expressed in terms of
$N$ nonnegative integers $(f_1,f_2,...,f_N)$ such that $f_1>
f_2>\cdots> f_N$. The ground state, $|0\rangle_{EV}$, is given by
$f_i=N-i$ and its energy will be $\sum_{n=0}^{N-1}(n+1/2)=N^2/2$.
It is more convenient to write the states in terms of nonnegative
integers which represent the excitations above the ground state
$(r_1,r_2,\cdots,r_N)$ where $r_i=f_i-(N-i)$ such that $r_1 \ge
r_2\ge\cdots \ge r_N\ge 0$. It was shown in \cite{Beren1} that the
above two bases are related by the Schur polynomial basis and the
two descriptions are identical. There is then a one-to-one
correspondence between the fermionic occupation indices $f_i$'s (or
$r_i$'s) and the bosonic harmonic oscillator $c_i$'s. The map
between the two is basically given by the bosonization of the 2d
fermion system. (This point has been discussed in a recent paper
\cite{Dhar}.) One may also use the difference of two successive
$r_i$'s for labeling the states, i.e. $(w_1,w_2,\cdots, w_N)$,
where $w_i=r_{i-1}-r_i$ for $1<i<N$ and $w_N=r_N$. In the Young
tableau notation the $w_i$ labeling corresponds to the Dynkin
labels, see Fig. \ref{Dynkin-figure}.

We now have all the ingredients to relate finite QHS to the 1/2
BPS sector of the gauge theory. This relation is most apparent
when we use the eigenvalue basis for the latter. In this basis, it
is immediately seen that the two descriptions are identical if and
only if $k=1$. It is an interesting and curious question whether
one can find deformations or perhaps other sectors of SYM which
could be described by fractional QHS i.e. with $k\ne1$. We will
address this question in the discussion section.

Restricting to $k=1$, consider for example the ground state
$|0\rangle_{QH,1}$. This is a circular droplet with radius
$R\sim\sqrt{N}$ and maps to the ground state of the gauge theory
or $|0\rangle_{EV}\equiv|0\rangle_{tr}$ which in turn corresponds
to the $AdS_5\times S^5$ background with $R_{AdS}^2\sim\sqrt{N}$.
The minimal excitation of the particle system amounts to exciting
the particle at the edge of the droplet to the state $|N\rangle$
which produces a circular hole state near the edge of the droplet.
In the eigenvalue basis this is the state $(1,0,...,0)$ and in the
trace basis it is produced by $\beta_1^\dagger|0\rangle_{tr}$. The
corresponding geometry is an $AdS_5\times S^5$ with a smeared
giant very close to the equator of $S^5$. As another example, a
hole in the origin of the droplet, $(1,1,...,1)$, is identical to
$\beta_N^\dagger|0\rangle_{tr}$ and corresponds to the largest
giant on the pole of $S^5$. A hole of unit area in a generic point
in the quantum Hall droplet, which is described by a coherent
state in the quantum Hall language, corresponds to a localized
giant graviton the radius squared of which  is proportional to its
distance from the edge of the droplet.

\section{Quasiholes in the $\cN=4$ SYM}\label{sec4}

In this section we attribute independent physical degrees of
freedom to the hole states of QHS by considering them as the
excitations of a dynamical field which we denote by $\Phi$. In
this way, the hole states are put on the same footing as the
particles and the duality between these states is promoted to a
dynamical one.

There are several motivations why one would like to treat the two
different excitations (particles Vs. holes) symmetrically. A
strong one comes from the gravity description of QHS in the LLL
with unit filling fraction $(\nu=1)$ which is given by the half
BPS geometries (LLM solutions) of type IIB SUGRA. (This point
has first been discussed in \cite{Bena-Smith}.) In
fact, for these geometries the above mentioned duality is enhanced
to a symmetry which manifests itself through the invariance of
solutions under the interchange of black and white boundary
conditions, accompanied by a change of orientation on the plane
$(x_1,x_2)$. This enhancement occurs because
for $\nu=1$ the statistics of particles, given by $\nu$, is the
same as that for the holes, given by $1/\nu$, and both are
fermions. Furthermore, the symmetry requires the same absolute
value for the charge units of particles and holes which holds only
when $\nu=1$.

{}From the quantum Hall physics side, adding new degrees of
freedom has been considered in order to accommodate quasiparticles
as well as quasiholes. The latter, as we reviewed in section
\ref{sec2.4}, can be achieved  via the edge state ({\it cf.}
\eqref{droplet+hole}). Quasiparticle states, however, were obtained
within the  model introduced in \cite{JackiwPi} in the context of
commutative Chern-Simons model and then in \cite{Bak, KMLee}
extended to the noncommutative Chern-Simons.

\subsection{Particle-Hole Lagrangian}\label{sec4.1}

To construct a symmetric model we introduce a new dynamical field
$\Phi$, in addition to $Z$, in the NCCS matrix theory and
interpret the excitations of this field as the hole states. We
choose $\Phi$ as an infinite dimensional complex matrix which is
in the adjoint of the gauge group. The important point is that
although we are dealing with an infinite four matrix theory, as we
will see, one can find solutions to this theory which represent
finite particle/hole systems. Therefore the gauge group dimension
in the particle/hole sector of the system comes up as a part of
the solution rather than as an input parameter.

Let us start with the NCCS matrix model Lagrangian we presented in section
\ref{sec2.3} %
\be\label{MatCS}%
 L_{CS}=-\frac{\pi\kappa}{\theta}
Tr\bigg(-\epsilon_{ij}X_i(\dot{X}_j+i[A_0,X_j])+2\theta A_0\bigg),
\ee %
where in terms of external magnetic $B$ field applied to QH
liquid in the unit of electric charge %
\be
2\pi\kappa=B\theta=\frac{1}{\nu}\ ,
\ee%
 and $\nu$ is the filling
fraction. If we expand the covariant position operator $X_i$
around the solutions of the equation of motion for $A_0$,
$[X_i,X_j]=i\theta\epsilon_{ij}$, in terms of the comoving
coordinates $y_i$'s as%
\be\label{XVs.y}%
X_i\equiv y_i+\theta\epsilon_{ij}A_j\ ,
\ee%
with $[y_i,y_j]=i\theta\epsilon_{ij}$, \eqref{MatCS} then reduces
to noncommutative Chern-Simons (NCCS) action in 2+1 dimensions
\cite{Susskind, Bak}. In this notation%
\be\label{Cov-Der}
\begin{split}
D_i\Phi &\equiv
\partial_i\Phi+i[A_i,\Phi]\cr
&=\frac{i}{\theta}\epsilon_{ij}[X_j,\Phi]\,,
\end{split}%
\ee%
where $\Phi$ is an arbitrary matrix which can also be thought of as a
field in the adjoint representation of NC $U(1)$.

To incorporate holes as independent dynamical degrees of freedom
we add a non-relativistic matter field $\Phi$ to the action
\eqref{MatCS}:
 \bea\label{p-h-sym-action}
L_{Z{\rm -}\Phi}&&=-\frac{\pi\kappa}{\theta}Tr\bigg(i(Z^\dagger
D_0Z-(D_0Z)^\dagger Z)+2\theta A_0\bigg)\cr &&\cr
&&+\frac{\pi\kappa}{\theta} Tr\bigg(i(\Phi^\dagger
D_0\Phi-(D_0\Phi)^\dagger \Phi) -\frac{1}{2m}D_i\Phi
(D_i\Phi)^\dagger-V(\Phi)\bigg)\label{lph},%
\eea%
where $Z=\frac{1}{\sqrt{2}}(X_1+i X_2)$ and $D_0 Z=\partial_0
Z+i[A_0,Z]$ (and similarly for $\Phi$). $m$ is the effective mass
for the $\Phi$ particle and we choose the potential $V(\Phi)$ to
be%
\be\label{potential}%
 V(\Phi)=-\frac {1}{2m\theta^2}
\bigg([\Phi,\Phi^\dagger]+\frac{\theta}{2}\bigg)^2. %
\ee%
The action \eqref{p-h-sym-action} is an extension of the
noncommutative version of Dunne-Jackiw-Pi-Trugenberger  model
\cite{JackiwPi}, discussed in \cite{KMLee}. Note that, unlike
\cite{Bak}, in our action $\Phi$ is in the adjoint ({\it cf.}
\eqref{Cov-Der}).

It is instructive to compare our $\Phi$ field with the edge state
$\Psi$ introduced in section \ref{sec2.4}. The $\Phi$ field being
an $N\times N$ matrix, rather than an $N$ vector, may be thought
of as a collection of $N$ number of edge states. In the classical
analysis of the Polykronachos' model \cite{Polychronakos} with one
edge state we could only describe a single droplet (which may have
a hole in it {\it cf.} \eqref{droplet+hole}). In order to describe
two droplets, or generically multi-droplets, such as multi
concentric rings in Fig \ref{Dynkin-figure}, {\it classically} we
need to introduce an edge state for each droplet. At quantum
level, i.e. in Calogero model, quantum fluctuations of a single
edge state, however, allows describing multi droplets. {}From the
APD symmetry viewpoint this can be understood noting that APD's at
classical level ($w_{\infty}$ transformations)  does not relate
configurations with different number of rings while at quantum
level ($W_{\infty+1}$ transformations) can tear the edge of a
droplet apart and hence relate a single droplet to two droplets
with the same area. In other words, the edge states corresponding
to different number of rings belong to topologically distinct
sectors of the gauge orbits of NC $U(1)$ in the NCCS theory or
$U(N)$ in the Chern-Simons Matrix theory \eqref{MatCS}. In this
viewpoint our model which {\it classically} contains the multi
edge state field $\Phi$, is an effective field theory description
of the quantum version of the Polychronakos model (the Calogero
model).

Before proceeding with the analysis of the action
\eqref{p-h-sym-action}, let us motivate the potential term
\eqref{potential}. As we have implicitly seen and would be
discussed further in the following section, the $\Phi$ (and
$\Phi^\dagger$) field corresponds to giant (anti-giant) gravitons
in the gravity picture and $V(\Phi)$ is representing the
giant-antigiant force and hence the potential \eqref{potential} is
a tachyonic potential corresponding to the open string tachyon in
the giant-antigiant system. To first order in $\alpha'$ this
potential can be obtained from the expansion of a Born-Infeld
action. (Note that potential \eqref{potential}, up to integrals of
total derivatives and a shift in zero point energy, is
proportional to $-Tr([\Phi,\Phi^\dagger]^2)$.)

 We start the analysis with
the equation of motion for $A_0$, the Gauss law constraint%
\be\label{G-L-Z-Phi}%
[Z,Z^\dagger]-[\Phi,\Phi^\dagger]=\theta\ .%
\ee%
Comparing \eqref{G-L-Z-Phi} with \eqref{x1x2-NC}, it is convenient
to choose $\theta=2\pi l_p^4$ and $m\theta=l_p$ and use the units
in which $Z$ and $\Phi$ are both measured in units of
$\sqrt{\theta}$, $\partial_0$ and $A_0$ in units of $1/l_p$. As
discussed earlier, in the 1/2 BPS sector of $\cN=4$ SYM we can only
realize a quantum Hall system with $\nu=1$ and hence we set $2\pi
\kappa=1$.

In the following we will show that this Lagrangain admits BPS
(solitonic) solutions. In order to do that we begin with the
Hamiltonian%
\be\label{Z-Phi-Hamiltonian}%
H=\frac{1}{2\theta}\bigg[\frac{1}{2m\theta^2}Tr([Z,Z^\dagger][\Phi,\Phi^\dagger]-2[Z^\dagger,\Phi^\dagger][Z,\Phi])+
Tr V(\Phi)\bigg]\ .
\ee%
Recalling the form of the potential \eqref{potential}, and using the
Gauss law constraint (\ref{G-L-Z-Phi}), it is readily seen that
\be\label{BPS}%
[Z,\Phi]=0%
\ee%
appears as the BPS condition. We should stress that to obtain a
BPS configuration \eqref{BPS} should be solved together with
\eqref{G-L-Z-Phi}.\footnote{As has been discussed in \cite{KMLee},
BPS solitonic equations \eqref{G-L-Z-Phi}, \eqref{BPS} can be
obtained from a noncommutative version of 2d chiral model. The
reduction from $(2+1)$ dim. to 2d may be understood noting that
BPS solitons are time independent. In \cite{KMLee} it has also
been argued that noncommutative 2d chiral model is solvable and
the moduli space of the solutions is trivial.} {}From the
equations of motion for $Z$, it is
inferred that for the static BPS configurations%
\be\label{A0-BPS}%
A_0=
\frac{1}{4m\theta^2}\left([\Phi,\Phi^\dagger]+\frac{\theta}{2}\right)=
\frac{1}{4m\theta^2}\left([Z,Z^\dagger]-\frac{\theta}{2}\right).%
\ee%
In the second equality we have used the Gauss law constraint
\eqref{G-L-Z-Phi}. For the BPS configurations the Hamiltonian
\eqref{Z-Phi-Hamiltonian} becomes a constant and%
\be\label{Z-Phidagger}%
[Z+\Phi^\dagger, (Z+\Phi^\dagger)^\dagger]=\theta\ . %
\ee%
The action evaluated on  a BPS configuration is
\[
L_{Z-\Phi}^{BPS}=-\frac{i}{2\theta}\bigg(Z^\dagger
\partial_0Z-(\partial_0Z)^\dagger Z -\Phi^\dagger
\partial_0\Phi+(\partial_0\Phi)^\dagger \Phi\bigg)\ .
\]

$\Phi$ fields which solve  the BPS equation \eqref{BPS}, if we
treat $\Phi$ as a function of $Z$ and $Z^\dagger$, are all
holomorphic functions of $Z$. This fact can directly be connected
with the holomorphicity of the chiral primary \opt s in $\cN=4$
SYM and/or the holomorphicity of the wavefunctions describing a
quantum Hall system in the lowest Landau level (i.e. Laughlin
wavefunction). In other words, the BPS configurations of the
action we have proposed are satisfying the same condition as the
(half) BPS configurations of the $\cN=4$ SYM. It is worth noting
that, as seen from \eqref{G-L-Z-Phi}, the classical solutions, BPS
or non-BPS, to the $Z$-$\Phi$ model are all infinite size
matrices.

\subsection{Static solitonic BPS solutions}\label{sec4.2}

In this section we find some static  classical solutions to the
BPS equations derived from Lagrangian (\ref{lph}). The BPS
configurations can be represented by the color coding: black
region for the particles, where $[Z,Z^\dagger]$ is non-zero, and
white region for quasiholes, where $[\Phi,\Phi^\dagger]$ is
non-vanishing.
\subsubsection*{\ \  {I. Vacuum}}%
As the first solution we consider the vacuum. This can be chosen
to be either the particle vacuum (black plane) or the hole one
(white plane). Let's choose the latter for which the solution is
simply
given by%
\be\label{vacuum-soln}%
Z=0\;,\;\;\;\;\;\Phi=\sum_{n=1}^\infty\sqrt{n\theta}|n\rangle\langle
n-1|\;.%
\ee%
 This solution for $\Phi$ can be interpreted as an in
finite number of concentric annular rings with unit area around
the origin. The inner and outer radii of the ring $n$,
${R_i}_{(n)}$ and ${R_o}_{(n)}$, can be determined by%
\be%
{R_o}^2_{(n)}=\theta\langle
n|\Phi_{n+1}^\dagger\Phi_{n+1}|n\rangle\;\;,{R_i}^2_{(n)}
=\theta\langle n|\Phi_{n}\Phi_{n}^\dagger|n\rangle\;, %
\ee%
where%
\be%
\Phi_n\equiv \sqrt{n\theta}|n\rangle\langle n-1|\;,\;\;\;\;\;n=1,2,\cdots%
\ee%
such that ${R_o}_{(n-1)}={R_i}_{(n)}=\sqrt{n\theta}$. The
Gauss law constraint implies that%
\be%
{R_o}^2_{(n)}=\theta+{R_i}^2_{(n)}\;. %
\ee%

\subsubsection*{\ \ {II. Black circular droplet}}

A black circular droplet is specified by the following solution%
\be\label{Disk-soln}%
\begin{split}
Z&=\sqrt{\theta}\sum_{n=1}^{N}\sqrt{n}|n-1 \rangle\langle n|\ ,\cr
\Phi&=\sqrt{\theta}\sum_{n=N+1}^{\infty}\sqrt{n}|n
\rangle\langle n-1|.%
\end{split}
\ee%
Here, the ring $N$ has its inner radius inside the particles and
its outer radius inside the holes. One can see that there is an
anomaly in $[Z,Z^\dagger]$ in the $|N\rangle\langle N|$ component
which is removed by $\Phi_{N+1}^\dagger\Phi_{N+1}$. Comparing this
to the finite matrix model of Polychronakos, it is obvious that
$\Phi_{N+1}^\dagger\Phi_{N+1}$ is behaving as $\Psi\Psi^\dagger$
where $\Psi$ is the edge state. Therefore the edge state for the
particle sector is provided by the hole degrees of freedom.
Alternatively, an edge state for the hole sector is provided by
the particle degrees of freedom through the term $Z_{N}^\dagger
Z_{N}$. This will be discussed further in section \ref{sec4.3}.

\begin{figure}[h]
\begin{center}
\epsfig{file=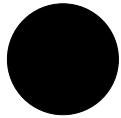,width=0.1\textwidth} \caption{A
circular droplet}
\end{center}
\end{figure}

\subsubsection*{\ \ {III. White rings inside a black circular droplet}}

\begin{figure}[h]
\begin{center}
\epsfig{file=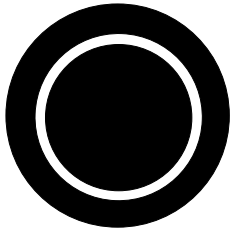,width=0.2\textwidth} \caption{A
circular droplet with a white ring. The area of the inner black
disk is $k_1$, the white ring $k_2$ and the outer black ring
$k_3$. This BPS solitonic configuration of our $Z$-$\Phi$ model
corresponds to $k_2$ smeared $S^5$-giant gravitons in a
$U(k_1+k_3)$ gauge theory or alternatively to $k_3$ smeared
AdS-giants in a $U(k_1)$ gauge theory.}\label{A-white-ring}
\end{center}
\end{figure}

As the next example we present the solution corresponding to a
black circular droplet with a concentric white ring inside as
depicted in Fig. \ref{A-white-ring}.

The solution is given by%
\bea\label{one-ring-soln}%
&&Z=\sqrt{\theta}\bigg(\sum_{n=1}^{k_1}\sqrt{n}|n-1\rangle\langle
n| +\sum_{n=q+1}^{q+k_3}\sqrt{n}|n-1 \rangle\langle n|\bigg)\ ,\cr
&&\\ &&
\Phi=\sqrt{\theta}\bigg(\sum_{n=k_1+1}^{q}\sqrt{n}|n
\rangle\langle n-1| +\sum_{n=q+k_3+1}^\infty\sqrt{n}|n
\rangle\langle n-1|\bigg)\ ,\nonumber%
\eea%
where $q=k_1+k_2$ and $k_1,k_2,k_3$ are respectively areas of the
inner black, white and outer black regions.

\subsubsection*{\ \ {IV. Plane-wave  solution}}

The plane-wave can be found either as an independent solution or
as a limit of the droplet. This amounts to pulling the top left
corner of the $Z$ and $\Phi$ matrices in the droplet solution to
infinity. It is readily seen that the resulting configuration, as
expected, describes a Fermi sea. Taking such a limit of the
droplet solution with white rings inside, will yield the ladder
configuration. One should note that in the latter case, the limit
is possible if the original solution describes very narrow black
and white rings near the edge of the droplet. This means that the
nonzero islands in $Z$ and $\Phi$ matrices which describe the
black and white stripes respectively, must be very small compared
to the dimension of the top left sub matrix of $Z$ which describes
the droplet.

\subsection{Closer connection to $N=4$ SYM and QH states}\label{sec4.3}
A 1/2 BPS \opt\ of $\cN=4$ $U(N)$ SYM is characterized by a couple
of quantum numbers. The first one which is not usually mentioned
is the size of matrices $N$. The second is R-charge $J$. With a
given $J$ and $N$, the number of traces (or number of
subdeterminants) in the operator constitute the other quantum
numbers. As reviewed in section \ref{sec3.2} and depicted in
Fig.\ref{Dynkin-figure} all this information can be summarized in
a Young tableau or configuration of concentric rings. It is
instructive to extract all these information from our $Z$ and
$\Phi$ matrices.

It is readily seen that the only non-zero  entries of the matrix
$Z^\dagger Z$ for a BPS configuration, in the harmonic oscillator
basis we have employed, are diagonal ones, where we have black
regions. Therefore, the number of non-zero elements of $Z^\dagger
Z$ is $N$.
In order to represent $N$ as  trace over a matrix, we construct a
regularized matrix inversion for any given matrix
$A$ %
\be\label{reg-inverse}%
 A^{-1}_{reg}\equiv \frac{1}{2} \lim_{\epsilon\to 0}\
 \left(\frac{1}{A+\epsilon {\bf 1}}+
\frac{1}{A-\epsilon{\bf 1}}\right) . %
\ee%
This definition of $A^{-1}$ reduces to the standard one when $A$
is invertible and has no zero eigenvalues. However, when $A$ has
zero eigenvalues is takes out the part with zero eigenvalues and
inverts the rest of the matrix, e.g.%
\[
A=diag(a_1,a_2,\cdots, a_k,0,0\cdots)\ \Rightarrow \ \
A^{-1}_{reg}=diag(\frac{1}{a_1},\frac{1}{a_2},\cdots,
\frac{1}{a_k},0,0\cdots)\ .
\]%

Using \eqref{reg-inverse} it is then easy to construct $N$ as%
\be\label{Nb-matrix}%
N_b= Tr\left((Z^\dagger Z) (Z^\dagger  Z)^{-1}_{reg}\right) \ .
\ee%
$\theta N_b$ is the area of the black region. Similarly one can
define 
$N_w\!=\!Tr\left((\Phi\Phi^\dagger)(\Phi\Phi^\dagger)^{-1}_{reg}\right)$ 
as the area of the white
region. In our case we have chosen $N_b$ to be finite while $N_w$
is infinite. Obviously both $N_b$ and $N_w$ are conserved quantum
numbers on the space of BPS solutions, as they are all static.

The quantum number $J$ is associated with the rotation symmetry in
the $Z$-plane. In other words, $J$ is the Neother charge for the
rotations $Z\to e^{i\phi}Z$, i.e. %
\be\label{J-matrix}%
J= Tr (Z^\dagger Z)-\frac{1}{2}N_b(N_b+1)\ ,%
\ee%
where we have subtracted off the ``zero point energy''
$\frac{1}{2}N_b(N_b+1)$ to match with the conventions of quantum
Hall system and the super Yang-Mills. One can also define $J_w=Tr
(\Phi \Phi^\dagger)$, which is again the Neother charge associated
with the rotations in the $\Phi$-plane. It is easy to see that for
the disk configuration \eqref{Disk-soln} $N_b=N$ and $J=0$ and for
the ring solution \eqref{one-ring-soln} $N_b=k_1+k_3$ and $J=k_2
k_3$.

In section \ref{sec2.2} we discussed that the density of the
quantum Hall fluid (in the Euler description) is
$\left(\theta\epsilon_{ij}\{X^i,X^j\}\right)^{-1}$
\eqref{rho-density}. Hence the inverse of $[Z,Z^\dagger]$ (or
$-[\Phi,\Phi^\dagger]$) should correspond to the particle (hole)
density%
\be\label{particle-hole-density}%
 \rho_{part}=[Z,Z^\dagger]^{-1}_{reg}\ \ ,\ \ \
 \rho_{hole}=[\Phi^\dagger,\Phi]^{-1}_{reg}\ .
\ee%
It is straightforward to see that for our concentric ring
solutions in the harmonic oscillator basis $[Z,Z^\dagger]$ is
diagonal and except for the boundaries of the black and white
regions, where we have color-changing, $\rho$ takes values zero or
one. This is as we expected and is made explicit in the LLM
construction. For example for the single ring configuration given
in \eqref{one-ring-soln}%
\be\label{rho-matrix}%
\begin{split}
\rho_{part}&\!=\!\!diag(\overbrace{1,1,\cdots,1}^{k_1},\frac{-1}{k_1},
\overbrace{0,0\cdots,
0}^{k_2-1},\frac{1}{k_1+k_2+1},\overbrace{1,\cdots,1}^{k_3-1},
\frac{-1}{k_1+k_2+k_3},0,\cdots)\ ,\\
\rho_{hole}&\!=\!\!diag(\underbrace{0,0,\cdots,0}_{k_1},\frac{1}{k_1+1},
\underbrace{1,1\cdots,1}_{k_2-1},\frac{-1}{k_1+k_2},\underbrace{
0,\cdots,0}_{k_3-1},\frac{1}{k_1+k_2+k_3+1},1,\cdots)\ .
\end{split} %
\ee%

Number of rings is then related to the number of non-integer
elements of $\rho$. Alternatively, one can show that%
\be\label{Ring-number}%
r\equiv \sharp {\rm Rings}=
\frac{1}{2}Tr \left((\rho_{part}[Z,Z^\dagger])-N_b-1\right) \ . %
\ee%
As we see all the information of the Young tableau can easily be
extracted from $Z^\dagger Z$ and $[Z,Z^\dagger]$ by methods
similar to those  mentioned above.

As discussed earlier, $\Phi$ matrices behave as (infinite)
collection of Polychronakos edge states $\Psi$. To make a closer
connection to the edge states, we note that diagonal elements of
$[\Phi^\dagger,\Phi]$ (or $[Z,Z^\dagger]$) are either zero or one
except on the edge of the rings. This happens in $2r+1$ points.
Hence $[\Phi,\Phi^\dagger]^2+[\Phi,\Phi^\dagger]$ is zero except
on the locus where we have color-changing.\footnote{A similar
observation has been made in \cite{Mandal}.  There, however, it
was proposed to take
$[\Phi,\Phi^\dagger]^2+[\Phi,\Phi^\dagger]=0$, ignoring the edge
effects.} Explicitly%
\be\label{edge-Phi}%
[\Phi,\Phi^\dagger]^2+[\Phi,\Phi^\dagger]=\sum_{n=1}^{2r+1}
\Psi_n\Psi^\dagger_n(\Psi_n\Psi^\dagger_n-1)\ ,%
\ee%
where $\Psi_n$ are the effective edge states we need to include in
the NC Chern-Simons Matrix theory to describe a configuration with
$r$ rings.\footnote{We would like to comment that in the
Polychronakos terminology in fact we do not need $2r+1$ edge
states and $r+1$ of them is enough; whenever we are moving from a
black region to white region we need an edge state. Or in terms of
our matrices that is the number of negative eigenvalues of the
$[Z,Z^\dagger]$ matrix.}

We should, however, emphasize an important difference between our
$\Phi$ field and the edge state(s) $\Psi$. As mentioned earlier,
our $\Phi$ field is a classical (effective) field theory
description of the quantized Polychronakos model. In the droplet
with a hole solution \eqref{droplet+hole} the size of the hole $q$
is not quantized. In our model, as it can be seen from
\eqref{one-ring-soln} (by setting $k_1=0$) the area of the hole is
quantized. In our $Z$-$\Phi$ symmetric model, unlike the
Polychronakos model,\footnote{It is notable that in the quantized
Polychronakos model, the Calogero model, the area of the holes $q$
is also quantized \cite{Polychronakos}.} the area of the white and
black regions, are {\it quantized} both in the {\it same} units,
$\theta$. The latter is a property of quantum Hall systems with
$\nu=1$. As the other manifestation of this ``quantization'' we
note that in our model the transition between black and white
regions does not occur suddenly. For example, as it is seen from
\eqref{rho-matrix} there are $2r+1$ points  where the change of
color happens. These are the places where $\rho_{part}$ and
$\rho_{hole}$ have overlap.

\subsection{More on symmetries of the $Z$-$\Phi$ model}\label{sec4.4}

As already discussed in the $\cN=4$ SYM a Young tableau may be
interpreted as a configuration of sphere giants or AdS (dual)
giants. On the other hand in principle, similar to the usual
D-brane case, we have the option of having anti-giants of either
kind. Of course in a 1/2 BPS sector with a given supersymmetry we
only see giants or anti-giants and not both. There are two
$\mathbb{Z}_2$ symmetries relating sphere giants and the AdS
giants and/or giants and anti-giants \cite{Alishah}. In our
$Z$-$\Phi$ model, in which giants and dual giants, respectively
denoted by $\Phi$ and $Z$, appear as independent degrees of
freedom one can realize both of the above $\mathbb{Z}_2$
symmetries. Both of these symmetries are essentially exchanging
the black and white regions.

The BPS equation \eqref{BPS} is manifestly invariant under the
exchange of $Z$ and $\Phi$ and the Gauss law constraint
\eqref{G-L-Z-Phi}, as well as \eqref{A0-BPS}, would remain
unchanged if we also send $\theta\to -\theta$. In other words,
\begin{subequations}\label{Z2one}
\begin{align}
Z &\longleftrightarrow \Phi\ ,\\ %
\theta&\longleftrightarrow -\theta\ ,
\end{align}
\end{subequations}
is a symmetry of BPS configurations. (\ref{Z2one}b) in the LLM
language means that besides the changing the black and white regions
one should also change the orientation of the $(x_1,x_2)$ plane
({\it cf.} \eqref{x1x2-NC}) \cite{Alishah}. The value of $A_0$
\eqref{A0-BPS}, noting the Gauss law constraint, remains unchanged
under the above $Z_2$ symmetry. In the quantum Hall language the
above is a particle$\leftrightarrow$quasihole symmetry.

The BPS configurations are invariant under another $\mathbb{Z}_2$
transformation which exchanges a giant with a dual anti-giant,
i.e.
\begin{subequations}\label{Z2two}
\begin{align}
Z &\longleftrightarrow \Phi^\dagger\ ,\\ %
t, A_0&\longleftrightarrow -t,-A_0\ .
\end{align}
\end{subequations}
In the above transformation we exchange black and white regions
without changing the $(x_1,x_2)$ plane orientation. In the quantum
Hall terminology \eqref{Z2two} which contains an inversion in time
is a particle/anti-quasihole exchange symmetry. Although our BPS
equations are $\mathbb{Z}_2$ invariant, their solutions are not
necessarily symmetric, as it is manifest in the black-white
diagrams. The only $\mathbb{Z}_2$ symmetric solution is when half
of the plane is filled by the black region, corresponding to the
plane-wave solution in the supergravity setup \cite{LLM}.

Compared to $\cN=4$ SYM the rank of the gauge group $N$, in our
model appears as a characteristic of the specific solutions rather
than a parameter in the $Z$-$\Phi$ Lagrangian. This observation
opens the way for extensions of AdS/CFT in which certain
quantities of $U(N)$ and $U(M)$ gauge theories are related. In
fact one can distinguish two different such extensions. The first
is inspired by Fig. \ref{A-white-ring} according which there is a
relation between giants of a $U(k_1+k_3)$ gauge theory and dual giants
of a $U(k_1)$ gauge theory.
 This is a very remarkable result, if it can be
extended to beyond the 1/2 BPS sector and to the full theory, as
one can then always use a ``dual'' picture in which the rank of
the gauge group is large and one can perform the 't Hooft
planar-nonplanar expansion.

The other such duality between gauge theories of different rank
may come from the above mentioned black/white exchange symmetry.
Consider the static solutions of our particle-hole symmetric
matrix model which describe a black droplet with circular white
rings inside.  \eqref{Z2one} exchanges the giants and dual giants,
if the $Z$'s are scalars of a $U(N+M)$ gauge theory, one then
expects that $\Phi$'s should become the scalars of a $U(M)$ theory
and vice-versa. To make the argument more tractable, we take the
matrices finite dimensional of size $N+M$ such that for each
solution, $[Z,Z^\dagger]$ has $N$ and $[\Phi,\Phi^\dagger]$ has
$M$ nonzero diagonal components.\footnote{Of course in our
$Z$-$\Phi$ model we are always dealing with infinite size
matrices. In order to keep both $N$ and $M$ finite, in the LLM
terminology, we should consider the case with  compact $(x_1,x_2)$
plane. This will bring further complications \cite{LLM} which
would be addressed in a future work \cite{Progress}.}
 Furthermore, we assume that $N\ll
M$. The ground state of such a system will thus be described by a
white 2-sphere with a very small black spot, on the north pole
say. This state is given by %
\bea%
[Z,Z^\dagger]_{ij}&=&\delta_{ij}\;,\;\;\;i,j=1,2,\cdots,
N\;,\;\;\;[Z,Z^\dagger]_{N+1N+1} =-N\;,\cr && \cr
[\Phi,\Phi^\dagger]_{ij}&=&\delta_{ij}\;,\;\;\;i,j=N+2,\cdots,
N+M\;,\;\;\; [\Phi,\Phi^\dagger]_{N+1N+1}=N+1\;.%
\eea%
 Excitations of the
system are represented by nonzero diagonal islands in the
commutators. Suppose an excitation which has an island of length
$n$ in $[Z,Z^\dagger]$ and an island of length $m$ in
$[\Phi,\Phi^\dagger]$. One can view this, as a 1/2 BPS excitation
of a $U(N)\ \cN=4$ SYM produced by ${({\beta_n^{(P)}}^\dagger)}^m$
acting on the gauge invariant vacuum or, equivalently, as such an
excitation of a $U(M)\ \cN=4$ SYM which is now produced by
${({\beta_m^{(H)}}^\dagger)}^n$ (where the superscript $P$ ($H$) on $\beta$
denotes Particle (Hole).) Remember that, as stated in
\ref{sec3.2},  ${\beta_n}^\dagger$ is the operator that produces
gauge invariant states in the trace basis by acting upon
$|0\rangle_{tr}$ \cite{Beren1, Nemani}.

In other words, the excitations of the black Fermi level can be
related to the hole states by going from trace to sub-determinant
basis in the $U(N)$ theory. On the other hand, the same particle
states can be related to the excitations of the white Fermi level
by going from sub-determinant to trace basis in the $U(M)$ theory.
As a result, the trace (sub-determinant) operators of the two
theories can be easily mapped to one another using the $Z$ and
$\Phi$ matrices.

\subsection{Stability analysis of the BPS configurations}\label{sec4.5}

So far we have stated the BPS equations of our $Z$-$\Phi$ model,
constructed and analyzed concentric ring solutions. In this
section we study classical stability of these BPS configurations.
To examine the stability let us start with the equations of motion
for $Z^\dagger$ and $\Phi^\dagger$:
\bea%
&&D_0Z-\frac{i}{4m\theta^2}[[\Phi,\Phi^\dagger],Z]+\frac{i}{2m\theta^2}[\Phi^\dagger,[Z,\Phi]]=0\,,\cr
&&\\
&&D_0\Phi+\frac{i}{4m\theta^2}[[Z,Z^\dagger],\Phi]+\frac{i}{2m\theta^2}[Z^\dagger,[Z,\Phi]]+
\frac{i}{2}\frac{\partial Tr V}{\partial\Phi^\dagger}=0\,. \nonumber%
\eea%
(The equations of motion for $Z$ and $\Phi$ are similar to these
equations obtained by getting a hermitian conjugate.) The above
should be solved together with the equation of motion for $A_0$,
the Gauss law constraint (\ref{G-L-Z-Phi}). In the gauge where
$A_0$ is given by \eqref{A0-BPS},  the equations of
motion simplify significantly to take the form%
\begin{subequations}\label{e.o.m}%
\begin{align}
\dot{\Phi}+\frac{i}{2m\theta^2}[Z^\dagger,[Z,\Phi]]&=0\,,\\
\dot{Z}+\frac{i}{2m\theta^2}[\Phi^\dagger,[Z,\Phi]]&=0\,,
\end{align}%
\end{subequations}%
where dot denotes the time derivative.

To address the classical stability of the BPS configurations first
we note that equations \eqref{e.o.m} are first order in time and
hence there is no non-trivial solution which  at $t=0$ is a BPS
configuration $\Phi(t=0)=\Phi_0$ and $Z(t=0)=Z_0$, where
$[Z_0,\Phi_0]=0$.
Next let us perturb the BPS solution as%
\be\label{perturb}%
\Phi=\Phi_0+\delta\Phi(t)\ ,\ \ \ Z=Z_0+\delta Z(t)%
\ee %
and let us suppose that there is no time that $\delta Z(t)$ and
$\delta \Phi(t)$ vanish simultaneously (otherwise the
perturbations would vanish for all $t$). Plugging the above into
\eqref{e.o.m} and \eqref{G-L-Z-Phi} expanding up to the first
order in perturbations we
obtain%
\be\label{variation-eom}%
\begin{split}
-i\frac{d}{d\tau}\delta
Z+[\Phi_0^\dagger,[Z_0,\delta\Phi]]+[\Phi_0^\dagger,[\delta
Z,\Phi_0]]&=0\ ,\cr
-i\frac{d}{d\tau}\delta\Phi\
+[Z_0^\dagger,[Z_0,\delta\Phi]]+[Z_0^\dagger,[\delta Z,\Phi_0]]&=0
,\cr
[Z_0^\dagger,\delta Z]+[Z_0,( \delta
Z)^\dagger]=[\Phi_0^\dagger,\delta\Phi]+[\Phi_0,&(
\delta\Phi)^\dagger]\ ,%
\end{split}%
\ee%
where $\tau={2m\theta^2} t$. In order to show the stability we
should argue that the perturbations do not grow in time. This can
be argued for noting the constants of motion. It can be directly
checked that any quantity of the form $Tr(A A^{-1}_{reg})$, which
is always integer valued, is (classically) conserved. (To see this
it is enough to note \eqref{reg-inverse}.) In particular the area of
the black region, $N$, and its second moment, $J$, number of rings
$r$ and  the Hamiltonian are conserved. Hence the perturbations
cannot grow in time.

As an example we solve \eqref{variation-eom} for perturbation
about the vacuum (the whole white solution) given through
\eqref{vacuum-soln}. For this background $Z_0=0$ and hence $\delta
\Phi=0$%
\be\label{vacuum-fluct}%
\frac{d}{d\tau} \delta Z-i[\Phi_0^\dagger,[\Phi_0,\delta Z]]=0\ ,%
\ee%
where $[\Phi_0,\Phi_0^\dagger]=-\theta$. It is readily seen that%
\be\label{fluct-soln}%
\delta Z=e^{-i\omega \tau} e^{k\Phi_0}\ e^{p \Phi_0^\dagger}%
\ee%
solves \eqref{vacuum-soln} with $\omega=\theta^2 k p$. Assuming
that $\delta Z$ is finite for large distances on the $x_1,x_2$
plane we have $k=-\bar p$ leading to $\omega=-\theta^2|k|^2$. Since
$\omega$ is real valued the perturbations does not grow in time.
The \eqref{fluct-soln}, as expected is a non-relativistic wave of
particles (black region) moving in e.g. $x_1$ direction. In
general we expect that a generic solution to \eqref{variation-eom}
to be a linear combination of the plane-waves of the form
\eqref{fluct-soln}. Alternatively one may construct a solution to
two dimensional wave equations with rotational symmetry, via the
Bessel functions. The latter would be more appropriate for
studying the fluctuations about the circular symmetric ring
solutions we considered in section \ref{sec4.2}.

Finally, from the above discussions it is inferred that
classically there is no transition between the BPS configurations.
That is, there is no solution which at two times $t_0$ and $t_1$,
$Z(t_0)\neq Z(t_1)$, is a BPS configuration. As discussed in
previous section $N$ and $J$ are both conserved charges and if
there is any transition, classically or quantum mechanically,
between various BPS solutions the initial and final states should
have the same $J$ and $N$. A quantum mechanical transition between
the rings, however, imply that the conservation of number of rings $r$, which
holds classically,  is violated quantum
mechanically. In fact it is not difficult to find paths (of course
not classical ones) which interpolate between the BPS
configurations which are not the dominant contributions compared
to the full $\cN=4$ SYM theory analysis \cite{Antal, SYM-giants}.

\section{Discussion and Outlook}\label{discussion-section}

\begin{figure}[h]
\begin{center}
\epsfig{file=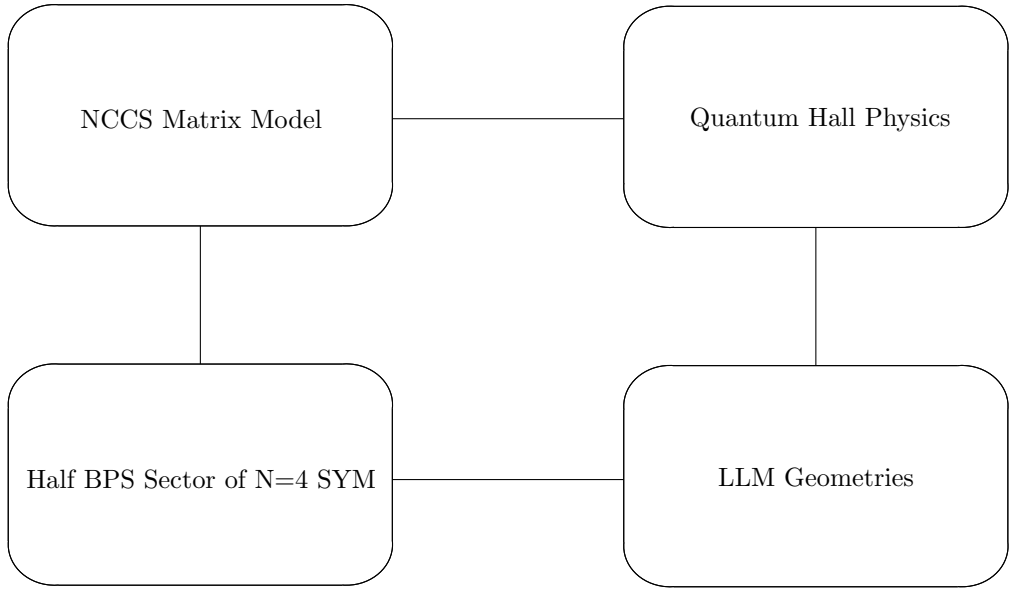}\caption{In this paper we have worked
out and discussed relation of all the four corners of the above
figure. The only case that we did not address here was a direct
connection of LLM geometries and noncommutative Chern-Simons
Matrix model. This may be done along the lines of minisuperspace
quantization method for the LLM geometries. This point has been
recently discussed in  \cite{minisuper}.}\label{Fig4}
\end{center}
\end{figure}

In this paper  we have studied, refined and clarified the
correspondence between the quantum Hall system and the $\cN=4$ SYM
in the 1/2 BPS sector and the LLM geometries, which was discussed
in \cite{LLM,Beren2}. In this viewpoint the 1/2 BPS sector of
$\cN=4$ SYM is equivalent to a QHS with filling fraction $\nu=1$.
We showed the equivalence of the two at the level of the actions
and the Hilbert spaces and/or the partition function of the
$\cN=4$ SYM in the 1/2 BPS sector.  We have shown that the (square
of) the Laughlin wavefunction with $\nu=1$ is nothing but the
partition function of the $\cN=4$ SYM in the 1/2 BPS sector. In
sum, we discussed and related four corners of the square depicted
in the Fig. \ref{Fig4}.

In the quantum Hall side the same machinery can be used for
$\nu<1$ systems, the fractional quantum Hall effect ($\nu>1$ has
problems with unitarity). We can now turn to the interesting
question of the relevance of FQHS in the context of AdS/CFT, an
issue which we advertised for in the paper. As mentioned before,
the physical states of a finite FQHS with $\nu=1/k$ can be found
as the states of a quantum Calogero model. The particles of this
system, obeying an enhanced Pauli exclusion principle, occupy the
energy states of a harmonic oscillator such that no two states
closer than $k$ are occupied. This results in the fact that on the
noncommutative plane where the particles live, the area quantum
for the particles is $k$ times bigger than that for the holes but
since we are probing the plane by the particles, and not with the
holes, $k+1$ possible values are found for the particle density.
Therefore one can take an equivalent but more convenient picture
of the system; think of particles as ordinary fermions but of
holes as obeying a ``reduced'' exclusion principle such that at
most $k$ number of these objects can be put on top of one another.
Such a ``reduction'' cannot happen for particles which is a result
of the upper bound on $\nu(\le1)$ and thus accumulation of
particles is not allowed.


One can make an analogous argument for the LLM geometries. These
solutions have two important features; noncommutativity of
$(z_1,z_2)$ and smoothness condition $(z_0=\pm1/2)$. The first one
arises upon the identification of the boundary plane with the
phase space of the dual fermionic system. Therefore, this feature
is originating from the quantum nature of the corresponding QHS.
In the geometric language, this noncommutativity is translated
through the statement that even a single giant graviton in the AdS
background produces a gravitational back reaction such that the
boundary configuration for the resulting geometry is depicted as a
white spot of minimal area inside a black droplet. This
noncommutativity of the $(x_1,x_2)$ plane is a result of quantum
gravity considerations, coming from the dual SYM picture, via
AdS/CFT. The smoothness condition of LLM geometries, on the other
hand, is a statement about the statistics of the particles and
holes in the dual QHS. As was found in \cite{Chronology}, in order
to exclude geometries with Closed Time-like Curves (CTC), one has
to impose an upper bound on the boundary value $z_0\le1$. In
parallel lines with the above arguments one can say that the dual
giants by which one probes the LLM backgrounds cannot be put on
top of one another, a ``stringy exclusion principle''
\cite{S-giants}, which is a result of the exclusion of CTC's in
the geometries $(z_0\le1)$. On the other hand, giant gravitons
also obey a ``stringy exclusion principle'' which is not reduced
because the regularity of solutions does not allow for
$-1/2<z_0<1/2$ and hence {\it there are no coincident giants or
dual giants  in the LLM setup}. Namely, LLM do not have such
solutions, because they only work with black and white regions. In
their setup this is a condition coming form smoothness of the
solutions. The latter have been made manifest in our $Z$-$\Phi$
symmetric model where $\Phi$ is associated with giants and $Z$
with dual giants.

In other words, Pauli exclusion principle which is there for both
quasiholes and particles in a quantum Hall system with $\nu=1$, is
a manifestation of the ``stringy exclusion principle'' coming out
of giants and dual giants considerations. (The stringy exlcusion principle in the context of 
supergravity has also been discussed in \cite{Bena-Smith}.)

As stated here the $\cN=4$ SYM in the 1/2 BPS sector is related to
quantum Hall system with $\nu=1$. One interesting open question is
whether the QHE/SYM correspondence can be pushed to $\nu<1$ where
we have the possibility of fractional quantum Hall effect and
anyons.  {}From our discussions it is inferred that such
correspondence, if it exists,  cannot happen on the 1/2 BPS sector
and we need to go beyond this sector.  One possibility, which is a
direct outcome of the discussions of the above paragraphs about
the fractional statistics of the quasiholes in the $\nu<1$ cases,
is to consider orbifolding in the $Z,Z^\dagger$ or
$\Phi,\Phi^\dagger$ plane. That is, considering a $S^5/Z_k$
orbifold of the $AdS_5\times S^5$ geometry where the orbifolding
is keeping an $SO(4)$ and is only acting on the $S^1$ in the
$Z,Z^\dagger$ plane. This orbifold has a fixed point at $Z=0$ and
is breaking all the supersymmetry. Upon orbifolding we are
identifying $k$ slices of the black disk, that is as if we have
$k$ particles of fractional statistics on top of each other. This
can happen if $\nu=1/k$. This proposal immediately tells why
$\nu^{-1}$ should be quantized. On the other hand, one can show
that the superstar solutions \cite{superstar}, which are 1/2 BPS
solutions of IIB supergravity with a naked singularity, are
effectively behaving like a quantum Hall system with $\nu<1$ (
more precisely, $\nu=\frac{1}{1+q}$ where $q$ is an integer
related to the $R$-charge of the solution.) The above proposal
then suggests that there should be a relation between the orbifold
singularity and the naked singularity of the superstar. Further
exploration of different aspects of this idea is postponed to
future works.

In the 1/2 BPS sector one can compute transition amplitudes
between giant gravitons. That is basically the computation of
three point function of three chiral primary operators carried out
in \cite{Antal}. This possibility has been ignored in the LLM
setup. In our $Z$-$\Phi$ model, as we argued our BPS solutions are
classically stable and as a result the number of giants is a
conserved quantity. (In the Calogero model there is no possibility
of such transitions between ring solutions, as states with
different number of rings (giants) are orthogonal eigenstates of
the Calogero Hamiltonian.) In the $Z$-$\Phi$ model, however, one
has the possibility of quantum tunneling between giants and hence
number of giants is not  conserved once the quantum
(non-perturbative instanton) effects are considered. This
expectation is compatible with the picture advocated in section
2.4 of \cite{Hedgehog}.

As another interesting extension one may try to push the  QHE/SYM
correspondence to beyond the 1/2 BPS sector. For example let us
consider a part of the 1/4 BPS sector only containing operator
made out of two of the three complex scalars, say $Z$ and $Y$. As
we argued in section \ref{sec3.1} one may think of the angular
momentum ($R$-charge) operator $J$ as the effective action for the
BPS sector, roughly that is $ Tr(Z \frac{\delta}{\delta Z})+Tr(Y
\frac{\delta}{\delta Y})$. This is very similar to a four
dimensional quantum Hall system. This is an open direction in need
of further analysis.

The last interesting open question we would like to briefly
discuss is the ``duality'' we alluded to at the end of section
\ref{sec4.4}. To argue for the ``duality'' between $U(N)$ and
$U(M+N)$ gauge theories we made the assumption that the total area
of the black and white regions is finite. In the LLM terminology,
that is the $(x_1,x_2)$ plane is compact and since $(x_1,x_2)$
plane is flat, our choices are limited to tori. (Note the footnote
at the end of page 7 in \cite{LLM} for a comment on this issue.)
 Since $(x_1,x_2)$ plane is  noncommutative, this torus should be a
 fuzzy torus. Exploring the possibility of compactifying $(x_1,x_2)$ plane
 from gravity and/or QH system side and its implication of that
 for the ``duality'' mentioned above is left for future works.
\vskip 1.0cm
\section*{ Acknowledgments}

We would like to thank  Jorge Russo for discussions and Vishnu
Jejjala for his collaboration at the early stages of this work.
A.E.M. would like to thank High Energy section of the Abuds Salam
ICTP where the last part of his work has been carried out. M.Sh-J
would also like to thank R. Asgari and Kimyeong Lee and Dongsu Bak
for helpful comments and discussions. O.S. would like to thank IPM
for the hospitality. The work of O.S. is supported by Ontario
Graduate Scholarship.


\end{document}